\documentclass[citeautoscript,showpacs,amssymb,twocolumn,prb,aps]{revtex4}
\usepackage{latexsym,graphicx,eufrak,amsmath}

\begin{document}
\title{Intermittency via Self-Similarity -- An Analytic Example}
\author{Mogens V. Melander and Bruce R. Fabijonas}
\affiliation{
Department of Mathematics,
Southern Methodist University,
Dallas, TX 75275}
\date{December 21, 2005; revised \today}
\begin{abstract}
Turbulence is known to show intermittency. That is, statistical properties 
vary with the length scale in a way not accounted for by statistical 
similarity where dimensionless 
ratios of moments are constant. Intermittency occurs even 
in the inertial range of isotropic turbulence, where physical intuition 
calls for a self-similar scale dependence. Perceived as a lack of 
overall scaling invariance, 
inertial range intermittency has become known as anomalous scaling. 
We present an analytic example demonstrating how anomalous
scaling and self-similarity in the form of global scaling 
invariance can coexist within the same statistics.
Whether we observe anomalous scaling or self-similarity 
depends on which variables we consider.  Our example
illustrates consequences of a symmetry, but is not meant
as an intermittency model.  
\end{abstract}
\maketitle

Intuitively,  fundamental insight is lacking if `anomalous' best
describes turbulence in its most ideal form.
Theoretically, 
turbulence is ideal when all simplifying circumstances are present. First, 
the fluid is incompressible and has constant properties--so the motion 
obeys the incompressible Navier-Stokes equations. Second, there are no 
boundaries and the turbulence is homogeneous and isotropic--so there is 
statistical independence of location and orientation. Third, steady large 
scale forcing puts the turbulence in equilibrium and makes its properties 
independent of time. Finally, the Reynolds number is high so that a wide 
range of scales separates forcing at the long lengths from dissipation at 
the short lengths. Inertial forces alone govern the physics in this 
``inertial range.'' Physical intuition dictates that the scale dependence in 
the inertial range must be self-similar--in some global way.
That is, full knowledge of the statistics at one scale should  
suffice to generate the statistics at all other scales.  

Kolmogorov \cite{K41} (K41) proposed a similarity like that 
in a Brownian motion 
where dimensionless ratios of moments are constant. Applied to the 
difference $\delta v(\ell )$ in 
a velocity component over a distance $\ell $, this so-called statistical 
similarity requires $\left\langle {\delta v^p} \right\rangle /\left\langle 
{\delta v^2} \right\rangle ^{p/2}$ to be independent of $\ell $
as suggested by dimensional analysis. Statistical 
similarity, however, does not agree with the evidence \cite{frisch}. 
The moment ratios 
are not constant. Equivalently, the probability density functions (pdf's) for 
$\delta v(\ell )$ at various $\ell $ do not collapse to one when plotted in 
units of standard deviations. Instead, one observes intermittency: 
fluctuations of many standard deviations become increasingly more likely at 
smaller $\ell $. Correspondingly, the tails of the pdfs flare out and become 
thicker. Although the moments are power laws, e.g., $\left\langle {\delta 
v^p} \right\rangle \propto \ell ^{\zeta _p }$, the exponent $\zeta _p $ 
varies non-linearly with the order $p$. In contrast, statistical similarity 
requires that $\zeta _p $ be linear. The name ``anomalous scaling'' alludes to 
historical frustrations in understanding why statistical similarity 
fails. Frisch \cite{frisch} presents an account of the 
history and theories for 
intermittency. Kolmogorov \cite{K62} addressed the 
issue with his 
log-normal model (K62). Although K62 has fallen out favor, it 
remains among the best-known intermittency models. 
To a large extent, it accounts 
for the discrepancies between the K41 predictions and experimental 
observations; see Ref.\,\onlinecite{pope00}, p.258.
However, new models, e.g. 
Refs.\,\onlinecite{She:Leveque:94}-\onlinecite{lun05}, 
describe observations better than K62.

Here, we show how self-similarity and anomalous scaling can 
coexist within the same statistics. The variables and functions we choose 
determine whether we observe self-similarity or only 
anomalous scaling. In particular, self-similarity emerges from considering the 
energy at scale $\ell $. To avoid lengthy mathematical derivations, we 
illustrate these ideas by an example. The example comes from a similarity 
theory we have developed for the inertial range \cite{mela:fabi:05}. 
This theory  puts the inertial range pdf for Navier-Stokes turbulence 
within the 
small class of functions generated by the inverse Mellin transform 
of $z^{-1}\exp \left( {\operatorname{sign}\left( {\beta -1} \right)z^\beta } \right)$. One 
of these functions has K62 scaling exponents, namely $\beta = 2$. 
We choose this 
function in our example 
for three reasons. First, the calculations work analytically and in a closed 
form independent of the theory \cite{mela:fabi:05}. Second, K62 is 
widely known and is in 
reasonable agreement with observations. Third, the concepts we  
emphasize are shared by all functions in the class: 
global scaling invariance (self-similarity) emerges 
through the energy at scale $\ell $ and coexists  with intermittency 
within the same statistics. 

Our example uses Mellin transforms extensively, so we review the 
needed properties. The Mellin transform 
\begin{equation}
\label{eq1}
\Phi (z)=\mathfrak{M}\left[ {\phi (x);z} \right]\equiv \int_0^\infty 
{x^{z-1}} \phi (x)\,dx 
\end{equation}
is useful for dealing with moments on the 
positive real axis. If known, $\Phi (z)$ provides all moments
including fractional orders. On the full 
axis, we have
\begin{equation}
\label{eq2}
\int_{-\infty }^\infty {x^p} \phi (x)dx=\mathfrak{M}\left[ {\phi (x);p+1} 
\right]+\left( {-1} \right)^p\mathfrak{M}\left[ {\phi (-x);p+1} \right].
\end{equation}
However, full range moments are defined only for integer orders. 
Consequently, we face ``Hausdorff's moment problem,'' where the integer 
moments do not uniquely identify a function; see Ref.\,\onlinecite{koerner}, 
p.21.  For example, 
many functions have the same integer moments as the log-normal. In contrast, 
the inverse Mellin transform is unique. The Mellin transform is closely 
related to the Fourier transform; see Ref.\,\onlinecite{sned95}, p.41. 
Each operational rule for 
the Fourier transform has a counterpart in the Mellin transform. We need the 
rule \cite{ober:74}
\begin{equation}
\label{eq3}
\mathfrak{M}\left[\phi \left( {\left( {x/a} \right)^{1/q}} \right);z\right]
   =a^zq\,\Phi (qz),
\quad
a,q>0\, .
\end{equation}
From tables of transforms \cite{ober:74} we need the entry
\begin{equation}
\label{eq4}
\mathfrak{M}\left[ {\tfrac{1}{2} \operatorname{erfc}\left( 
  {\tfrac{1}{2}\ln r} \right);z} 
\right]=z^{-1}e^{z^2} ,
\end{equation}
where $\operatorname{erfc}(t)\equiv 2\pi^{-1/2}\int_t^\infty 
  {e^{-\tau ^2}d\tau } $
is the complementary error function.

For our example, we design a function $F(r,\ell)$ with global 
scaling invariance in $\rho = \ln r$, but with anomalous scaling in 
$r$.  The self-similarity $F(r,\ell) = f((\rho-\mu(\ell))/\sigma(\ell))$
expresses our scaling invariance as an affine transformation 
on the $\rho-$axis.  That is, $\mu$ and $\sigma$ reset the origin
and the unit on the axis in the same way as we traditionally use 
mean and standard deviation.  We shall call the invariance under 
this affine transformation `normal scaling' to draw the distinction
from `anomalous scaling.'  Anomalous scaling requires the $p$'th moment of 
$F(r,\ell )$ to be a power law $C_p \ell ^{\xi _p }$ where $\xi _p $ depends 
non-linearly on $p$. Only specific combinations of $f(\rho)$, $\mu(\ell)$, 
and $\sigma(\ell)$ permit $F(r,\ell)$ to have anomalous scaling in $r$.
$f(\rho )=\tfrac{1}{2}\operatorname{erfc}\left( \tfrac{1}{2}{\rho } \right)$
is one such function as we shall now show. By 
construction, we have
\begin{equation}
\label{eq5}
F(r,\ell )=f\left( {\frac{\ln r-\mu (\ell )}{\sigma (\ell )}} 
\right)=f\left( {\ln \left( {\frac{r}{e^\mu }} \right)^{1/\sigma }} \right).
\end{equation}
Using (\ref{eq3}) we compute the moments:
\begin{multline}
\label{eq6}
\int_0^\infty {r^pF(r,\ell )\,dr} =\mathfrak{M}\left[ {F(r,\ell );p+1} 
\right] \\ =\sigma e^{(p+1)\mu }\mathfrak{M}\left[ {f(\ln r);\left( {p+1} 
\right)\sigma } \right]=\frac{e^{(p+1)\mu +(p+1)^2\sigma ^2}}{p+1}.
\end{multline}
For $\ell \le 1$ (small scales), we choose $\mu (\ell )=-\alpha \ln \ell $ 
and $\sigma (\ell )=\gamma \sqrt {-\ln \ell } $ so as to obtain
the power laws 
\begin{equation}
\label{eq7}
\int_0^\infty {r^pF(r,\ell )dr=} \frac{\ell ^{-(p+1)\alpha -(p+1)^2\gamma 
^2}}{p+1}.
\end{equation}
So, the scaling in $r$ is anomalous except when $\gamma =0.$ 

Let us establish a connection with turbulence. 
Because of incompressibility, the velocity is 
divergence-free and consequently has only two independent components. 
Thus, two random variables describe isotropic turbulence at each scale. 
One choice is longitudinal and 
transverse velocity increments, i.e., $\delta v_\parallel ( \ell 
)$ and $\delta v_\bot ( \ell )$. Another is left- and 
right-handed velocity components obtained via the complex helical waves 
decomposition \cite{lesieur}. Theoretically, we want the left-right symmetry
of the Navier-Stokes equations reflected in the variables.  Thus, their 
squares summed with equal weight should yield the energy in scale $\ell $. 
That then provides a Cartesian description. 
To emphasize this point, we call the 
two random variables $X$ and $Y$. In this regard, left- and right-handed 
amplitudes are good variables \cite{lesieur}, 
but $\delta v_\parallel ( \ell )$ 
and $\delta v_\bot ( \ell )$ are not; see 
Ref.\,\onlinecite{hinze87}, p.208. Let 
$\mathcal{X}(x,\ell )$ be the pdf for $X$, 
i.e., $\mathcal{X}(x)\,dx=\Pr\left\{ {x<X<x+dx} \right\}$. 
Similarly, let $\mathcal{Y}(y,\ell )$ be the pdf for $Y$, 
and $\mathcal{J}(x,y,\ell )$ the joint 
pdf for $X$ and $Y$. Because $\mathcal{J}(x,y,\ell )\to 0$ 
as $x^2+y^2\to \infty 
$, the graph of $\mathcal{J}$ has the shape of a mountain 
centered near $(0,0)$. Thus, 
we use polar coordinates $\left( {x,y} \right)=r\left( {\cos \theta ,\sin 
\theta } \right)$ and we obtain an azimuthal decomposition 
\begin{equation}
\label{eq8}
\mathcal{J}(x,y)=P(r)+\cos (\theta -\theta _1 )P_1 (r)
  +\cos (2(\theta -\theta _2 ))P_2(r)+\cdots
\end{equation}
where $\theta _1 ,\theta _2 ,...$ are phase-constants. We focus on the 
axisymmetric contribution to $\mathcal{J}$. Returning 
to our example, we construct $\mathcal{J}$ from $F(r,\ell )$ :
\begin{equation}
\label{eq9}
\mathcal{J}(x,y,\ell )=P(r,\ell )=C(\ell )F(r,\ell ),
\end{equation}
where the factor
$C(\ell )=1/\left( {2\pi \mathfrak{M}\left[ {F(r,\ell );2} \right]} 
\right)=\pi ^{-1}e^{-2\mu -4\sigma ^2}$ gives unit volume under the graph and 
ensures $\mathcal{J}$ is a pdf. 

\begin{figure}
\includegraphics[scale=0.38]{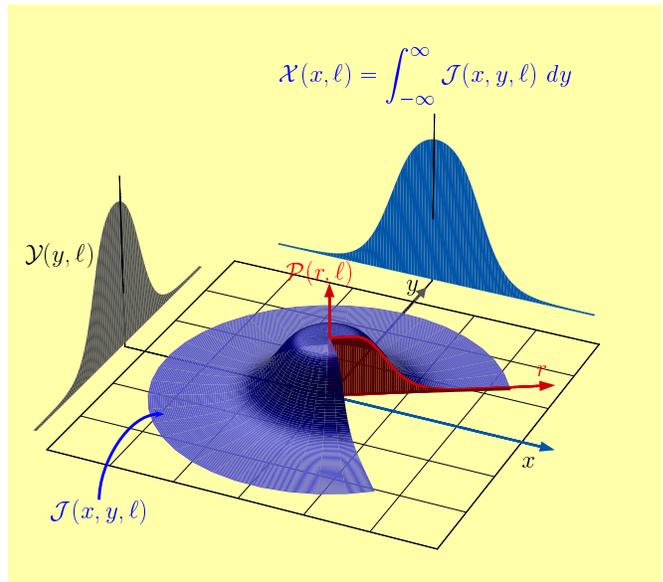}
\caption{(Color online) The joint pdf $\mathcal{J}(x,y,\ell )$ is 
shown for $\ell =0.01$ together with 
the one dimensional pdf $\mathcal{X}(x,\ell )$ and $\mathcal{Y}(y,\ell )$. 
The radial profile 
$P(r,\ell )$ is self-similar (Fig. 2), but $\mathcal{X}(x,\ell )$ shows 
anomalous scaling (Fig. 3). }
\end{figure}
With normal scaling built into $\mathcal{J}$, consider the situation 
graphically. Figure 1 shows $\mathcal{J}(x,y,\ell )$ for $\ell =0.01$. 
As we vary $\ell 
$ the graph changes. $\mu (\ell )$ controls the half-width-radius of the 
``pie,'' while $\sigma (\ell )$ sets the slope there. $\mu (\ell )$ and 
$\sigma (\ell )$ do not change proportionally, so statistical similarity is 
impossible. In the limit $\ell \to 1^-$, the graph looks like a disk, but 
as $\ell \to 0^+$ a sharp peak develops at $(0,0)$. Figure 2a shows the 
normalized radial profile for various $\ell $. For small $\ell $, the curves 
cluster near the vertical axis and are impossible to distinguish. The remedy 
is a logarithmic axis (Fig. 2b). On the logarithmic axis, the curves look 
alike so we align the midpoints (Fig. 2c). Then horizontal scaling 
(i.e. multiplying the abscissa by $1/\sigma$)
collapses all curves onto $f(\rho )=\tfrac{1}{2}\operatorname{erfc}\left(\tfrac{1}{2}{\rho } \right)$. 
\begin{figure}
\includegraphics[scale=0.40]{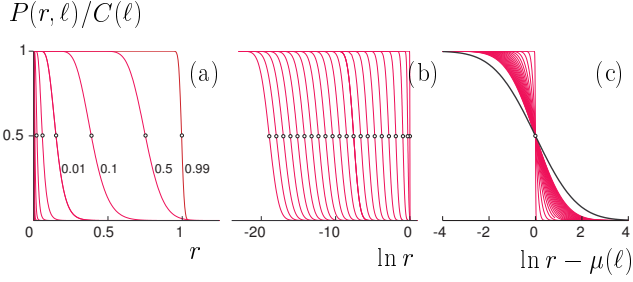}
\caption{(Color online) (a) The normalized radial profile 
$P(r,\ell)/C(\ell) = F(r,\ell)$ for $\ell 
=0.99,0.5,0.1,0.01, \ldots,10^{-20}$; 
(b) Same graphs on a logarithmic abscissa $\rho = \ln r$; 
(c) Collapse by normal scaling $\rho \to (\rho-\mu(\ell))/\sigma(\ell)$ 
to $\tfrac{1}{2}\operatorname{erfc}(\tfrac{1}{2}\rho)$
(the heavy black graph).}
\end{figure}

In contrast, the traditional analysis of the same statistics shows anomalous 
scaling. Two steps are involved. First, we find scaling exponents $\zeta _p 
$ for the structure functions (i.e., moments) of a single random variable 
such as $X$. Second, we plot the corresponding pdf, 
$\mathcal{X}(x,\ell )$, for 
various $\ell $ with the abscissa in units of $\left\langle {X^2} 
\right\rangle ^{1/2}$ and the ordinate scaled to give unit area. 

To obtain $\zeta _p $, we calculate the moments of $\mathcal{X}(x,\ell )$ by 
integrating $\mathcal{J}(x,y,\ell )$ over $y$. Using (\ref{eq7}) and (\ref{eq9}) we have 
\begin{multline}
\int_0^\infty {x^p} \mathcal{X}(x,\ell )\,dx=\int_0^\infty {x^p} \int_{-\infty }^\infty 
{P(r,\ell )} \,dy\,dx \\ =\int_{-\pi /2}^{\pi /2} {\int_{-\infty }^\infty {\left( 
{r\cos \theta } \right)^pP\left( {r,\ell } \right)} r\,dr\,d\theta } 
\\ \label{eq10}
=2\int_0^{\pi /2} {\cos ^p\theta \,d\theta \int_0^\infty {r^{p+1}P\left( 
{r,\ell } \right)} \,dr} \\ =K_p  \mathfrak{M}\left[ {P(r,\ell );p+2} \right]=
K_p \,\ell ^{-\alpha p-(p^2+4p)\gamma ^2}
\end{multline}
where $p>-1$ and 
$K_p =\sqrt \pi \Gamma \left( {\frac{p}{2}+\frac{1}{2}} \right)/\Gamma 
\left( {\frac{p}{2}+1} \right)$ with $\Gamma (z)$ 
being the usual gamma function. 
(Full range moments follow from (\ref{eq2}).) 
Note the $\theta-$integral in \eqref{eq10} converges only when 
$p>-1$, whereas the $r$ integral converges for $p>-2$.  Thus, 
by considering $X$ rather than $\sqrt{X^2+Y^2}$ we loose all 
exponents in the range $-2<p\leq -1$.
Clearly, $\zeta _p =-\alpha 
p-(p^2+4p)\gamma ^2$. So $\zeta _0 =0$ and $\zeta _3 =-3\alpha -21\gamma 
^2$. Choosing $\alpha =-\frac{1}{3}-7\gamma ^2$ and $\gamma =1/\sqrt {90} $ 
yields $\zeta _3 =1$ and $\zeta _6 =1.8$. By design, our $\zeta _p $, shown 
in the inset in Fig. 3, have classical K62-values \cite{frisch}. 
\begin{figure}
\includegraphics[scale=0.55]{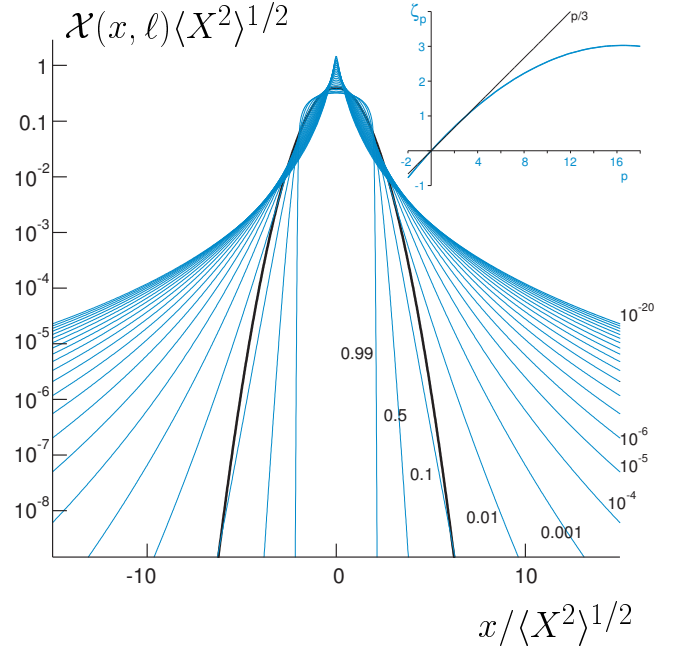}
\caption{(Color online) (Inset) K62 Scaling exponents; 
(Main figure) $\mathcal{X}(x,\ell )$ obtained numerically from 
(\ref{eq12}) for the same values of 
$\ell $ as in Fig. 2.}
\end{figure}

Figure 3 shows the graphs of $\mathcal{X}(x,\ell )$ for the same values of $\ell $ as 
in Fig. 2. This time, the curves do not collapse to one, but exhibit the 
usual flaring tails associated with anomalous scaling. To reveal asymptotic 
properties, we rewrite $\mathcal{X}(x,\ell )$ analytically. For $x>0$, 
\begin{multline}
\label{eq11}
\mathcal{X}(x,\ell )=\int_{-\infty }^\infty {\mathcal{J}(x,y,\ell )\,dy=} \int_{-\infty }^\infty 
{P\left( {r,\ell } \right)\,dy } \\ = \int_{-\pi /2}^{\pi /2} {P\left( {x\sec 
\theta ,\ell } \right)x\sec ^2\theta \,d\theta }  \\ 
=\frac{2x}{\sqrt \pi }C(\ell )\int_0^{\pi /2} 
{\int_{\frac{1}{2\sigma }\ln \left( {e^{-\mu }x\sec \theta } 
\right)}^\infty {e^{-t^2}\sec ^2\theta\,dt \,d\theta } }\\ 
=\frac{xC}{\sigma \sqrt 
\pi }\int\limits_0^{\pi /2} {\exp \left( {-\frac{\ln ^2\left( {e^{-\mu 
}x\sec \theta } \right)}{4\sigma ^2}} \right)\tan ^2\theta \,d\theta } ,
\end{multline}
where the last step requires integration by parts. Upon the substitution $\eta 
=\ln \sec \theta $, the integral becomes numerically friendly 
with nice asymptotic properties:
\begin{equation}
\label{eq12}
\mathcal{X}(x,\ell )=\frac{Ce^{\mu +\sigma ^2}}{\sigma \sqrt \pi }\int_0^\infty {\sqrt 
{1-e^{-2\eta }} } \exp \Big( {-\Big( {\frac{\eta -\eta _c }{2\sigma }} 
\Big)^2} \Big)d\eta 
\end{equation}
where $\eta _c =2\sigma ^2-\ln x+\mu $. 
For fixed $\ell $, we have $\eta _c \to \infty $ as 
$x\to 0^+$ and the integrand reduces to a Gaussian so that
\begin{equation}
\label{eq13}
\mathcal{X}(0,\ell )=\frac{Ce^{\sigma ^2+\mu }}{\sigma \sqrt \pi }\sqrt {4\pi \sigma 
^2} =2\pi ^{-1}e^{-3\sigma ^2-\mu }.
\end{equation}
With $\left\langle {X^2} \right\rangle ^{1/2}=\tfrac{1}{2}e^{6\sigma ^2+\mu }$, we 
have $\left\langle {X^2} \right\rangle ^{1/2}\mathcal{X}(0,\ell )=\pi^{-1}e^{3\sigma ^2} 
=\pi^{-1}\ell ^{-1/30} \to \infty $ as $\ell \to 0^+$. Thus, the peak in 
Fig. 3 rises without bound as $\ell \to 0^+$. Again for fixed $\ell $, we 
have $\eta _c \to -\infty $ as $x\to \infty $ so that the essential 
contribution to the integral \eqref{eq12} 
comes from a small interval to the right of 
$x=0$. Asymptotic analysis yields:
\begin{multline*}
\mathcal{X} \sim \frac{2x\sigma^2 }{\pi }e^{-4\sigma ^2-2\mu }\exp 
\left (-\frac{\ln ^2(xe^{-\mu })}{4\sigma ^2}\right )\ln ^{-3/2}(xe^{-\mu }) \\ =
\frac{2\gamma^2 |\ln \ell| }{\pi }\ell^{2\alpha+4\gamma^2}
x \ln ^{-3/2}(x/\ell ^\alpha )\exp 
\left( {-\frac{\ln ^2(x/\ell ^\alpha )}{4\gamma ^2\left| {\ln \ell } 
\right|}} \right) \label{eq14}
\end{multline*}
as $x\to \infty $. 
The decay of $\mathcal{X}(x,\ell )$ is essentially log-normal. 

We emphasize that our example is not intended as a model of 
intermittency, but rather as an illustration of the symmetry
expressed by the global invariance of $F(r,\ell) = P(r,\ell)/C(\ell)$
under the affine transformation 
$\rho(r)=\ln r\rightarrow(\rho-\mu(\ell))/\sigma(\ell)$.  Our example 
is not a model because we make no modeling approximations.  Rather, 
we suggest that equilibrium turbulence has precisely this symmetry.
Considering Navier-Stokes equations with steady forcing
as a dynamical system, equilibrium turbulence
represents an attractor in phase space.  Our suggestion 
is that  Navier-Stokes
equations restricted to the living space of the attractor obey our symmetry. 
Our suggestion is supported by shell model calculations \cite{mela:fabi:05}
and recently by low resolution DNS in a periodic box.
Theoretically, we can show\cite{mela:fabi:05} that the symmetry 
exists together with anomalous scaling in $r$ only when 
$\mathfrak{M}[f(\ln r);z] = z^{-1}\exp(\operatorname{sign}(\beta-1)z^\beta)$.  
Our preliminary DNS show $\beta \approx 1.4$, 
significantly less that two.  Thus, there is no point in attempting to 
fix and resurrect K62.

\begin{figure}
\includegraphics[scale=0.42]{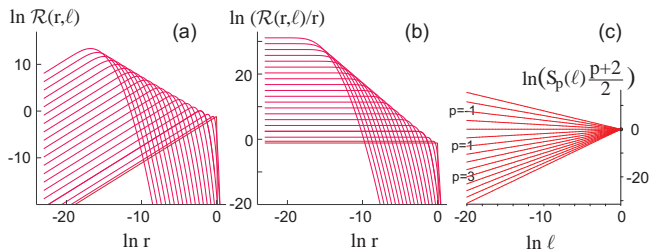}
\caption{(Color online) 
(a) The pdf for the amplitude $\sqrt{X^2+Y^2}$ at scale $\ell $ 
with the same $\ell$ values as in Figs. 2 and 3; 
(b) $\mathcal{R}(r,\ell )/r=2\pi 
P(r,\ell )$ is self-similar and collapses by normal scaling; (c) Structure 
functions 
$S_p (\ell )=2\ell ^{\zeta _p }/( {p+2} )$
plotted to reveal the virtual origin for the inertial range.}
\end{figure}
Using our example, we illustrate how our global scaling 
invariance appears in analysis of data with $J(x,y,\ell)$ 
statistics.  With $X^2+Y^2$  being a random variable for the 
energy at scale $\ell$, we construct the pdf for the 
velocity magnitude as $\mathcal{R}(r,\ell)\,dr = \operatorname{Pr}\{ 
r < \sqrt{X^2+Y^2} < r+dr\}$.  A simple and robust technique
for that purpose is kernel density estimation \cite{sil90}.  
Plotting the pdf on log-log scales for various $\ell$, 
we obtain Fig. 4(a).  All curves have an asymptote with 
slope one.  This is the signature of a monopole in 
polar coordinates, i.e. $0 < P(0,\ell) < \infty$ implies
$ \mathcal{R}(r,\ell)\,dr \approx 2\pi P(0,\ell)\,r\,dr$ as 
$r\to 0^+$.  We make the asymptote horizontal and thus immune to 
abscissa scaling by considering the function $\mathcal{R}(r,\ell)/r$,
which is $2\pi P(r,\ell)$ so that the curves in Fig. 4b collapse
to one by normal scaling.  In addition, our global scaling invariance
also shows up in $\mathcal{S}_p(\ell) \equiv
\langle (\sqrt{X^2+Y^2})^p\rangle$.  Using \eqref{eq5} and 
\eqref{eq9}, we have
$S_p (\ell )=2\ell ^{\zeta _p }/( {p+2} )$.
When plotted as in Fig. 4c, all lines converge on a focal point. It is at 
$(0,0)$ because of the way we have normalized our example. The focus marks 
an infrared virtual origin for the inertial range. The focal abscissa is the 
intrinsic length scale for inertial range. The ordinate is proportional to 
the dissipation as in Kolmogorov's four-fifths law, but we do not yet know 
the proportionality constant.

Consider the juxtaposition of Figs. 2 and 3, which represent 
the same statistics, $\mathcal{J}(x,y,\ell )$.  
In Fig. 2 all curves collapse to 
one by normal scaling, while in Fig. 3 anomalous scaling prevents them 
from doing so. Our example is analytic and exact--no modeling or numerical 
issues blur the juxtaposition. Our example shows directly that intermittency 
is not at odds with global scaling invariance. 
Normal scaling on the logarithmic axis 
accommodates two independent parameters $\mu \left( \ell \right)$ and $\sigma 
\left( \ell \right)$; see Fig. 2(b)-(c). In contrast, statistical 
similarity allows only one and fails for that reason.

\end{document}